\documentclass[proof]{pasj01}
\input{pasjsup.sty}
\usepackage{lscape}
\usepackage{color}

\def\commenta{$^*$}
\def\commentb{$^\dagger$}
\def\commentc{$^\ddagger$}
\def\commentd{$^\S$}
\def\commente{$^\|$}
\def\commentf{$^\#$}
\def\commentg{$^{**}$}
\def\commenth{$^{\dagger\dagger}$}
\def\commenti{$^{\ddagger\ddagger}$}

\newcounter{author}
\begin{document}
\SetRunningHead{K. Namekata et al.}{Validation a scaling law of solar and stellar flares}
\Received{201X/XX/XX}%{yyyy/mm/dd}
\Accepted{201X/XX/XX}%{yyyy/mm/dd}

\title{Validation of a Scaling Law for the Coronal Magnetic Field Strengths and Loop Lengths of Solar and Stellar Flares}
\author{Kosuke \textsc{Namekata},\altaffilmark{1,*} 
Takahito \textsc{Sakaue},\altaffilmark{1}
Kyoko \textsc{Watanabe},\altaffilmark{2}
Ayumi \textsc{Asai}\altaffilmark{3}
and Kazunari \textsc{Shibata}\altaffilmark{3}
}
\altaffiltext{1}{Department of Astronomy, Kyoto University, Kitashirakawa-Oiwake-cho, Sakyo-ku, Kyoto 606-8502, JAPAN}
\altaffiltext{2}{National Defense Academy of Japan, 1-10-20 Hashirimizu, Yokosuka, 239-8686, JAPAN}
\altaffiltext{3}{Kwasan and Hida Observatories, Kyoto University, Yamashina, Kyoto, 607-8471, JAPAN}
\altaffiltext{*}{namekata@kwasan.kyoto-u.ac.jp}
\maketitle

\begin{abstract}
\authorcite{1999ApJ...526L..49S} (\yearcite{1999ApJ...526L..49S}, \yearcite{2002ApJ...577..422S}) proposed a method of estimating the coronal magnetic field strengths ($B$) and magnetic loop lengths ($L$) of solar and stellar flares, on the basis of magnetohydrodynamic simulations of the magnetic reconnection model. 
Using the scaling law provided by \authorcite{1999ApJ...526L..49S} (\yearcite{1999ApJ...526L..49S}, \yearcite{2002ApJ...577..422S}), $B$ and $L$ are obtained as functions of the emission measure ($EM=n^2L^3$) and temperature ($T$) at the flare peak.
Here, $n$ is the coronal electron density of the flares. 
This scaling law enables the estimation of $B$ and $L$ for unresolved stellar flares from the observable physical quantities $EM$ and $T$, which is helpful for studying stellar surface activities.  
To apply this scaling law to stellar flares, we discuss its validity for spatially resolved solar flares. 
$EM$ and $T$ were calculated from \textit{GOES} soft X-ray flux data, and $B$ and $L$ are theoretically estimated using the scaling law. 
For the same flare events, $B$ and $L$ were also observationally estimated with images taken by \textit{Solar Dynamics Observatory} (\textit{SDO})/ Helioseismic and Magnetic Imager (HMI) Magnetogram and Atmospheric Imaging Assembly (AIA) 94{\AA} pass band. 
As expected, a positive correlation was found between the theoretically and observationally estimated values. 
We interpret this result as indirect evidence that flares are caused by magnetic reconnection.
Moreover, this analysis makes us confident in the validity of applying this scaling law to stellar flares as well as solar flares.
\end{abstract}

\section{Introduction}\label{sec:int}
%フレアとは
Flares are the most energetic phenomena on stellar surface. 
It is widely believed that, during flares, magnetic energy is converted into kinetic and thermal energy through magnetic reconnection in the corona. 
Fast reconnection creates reconnection jets upwards and downwards from the reconnection region, and superhot components are formed on the top of the magnetic loop due to jet collisions (see e.g., \cite{1981sfmh.book.....P,2011LRSP....8....6S,2016ASSL..427..373S}). The heat conduction and non-thermal particles along the magnetic field transport the thermal and non-thermal energy from the reconnection region to the foot point, and chromospheric evaporation then occurs \citep{1968ApJ...153L..59N,1974SoPh...34..323H}. Flares are observed with various wavelength ranges \citep{1974IAUS...57..105K}, and several aspects of the emissions can be explained by the above flare model. 

%スケーリングの紹介
According to \citet{1995ApJ...451L..79F}, \citet{1995PASJ...47..251S}, and \citet{1997PASJ...49..115Y}, there is a universal correlation between the temperature ($T$) and emission measure ($EM$) among not only solar flares but also stellar flares. \authorcite{1999ApJ...526L..49S} (\yearcite{1999ApJ...526L..49S}, \yearcite{2002ApJ...577..422S}) suggested a theoretical scaling law describing this correlation. Their theory is based on two-dimensional magnetohydrodynamic simulations of magnetic reconnection with heat conduction and chromospheric evaporation from a previous study \citep{1998ApJ...494L.113Y}, which showed that the reconnection heating ($B^2v_A/4\pi L$) is roughly balanced with the conduction cooling ($\kappa_0T^{7/2}/2L^2$) at the flare peak. They also assumed that the gas pressure ($p$) is comparable to the magnetic pressure ($B^2/8\pi$) in the evaporated plasma. Here, $v_A$ is the Alfv\'en speed and $\kappa_0$ is Spitzer's thermal conductivity. They then deduced the following scaling law:
\begin{eqnarray}
B_{\rm theor}=50 \left( \frac{EM}{10^{48}\rm{cm}^{-3}} \right) ^{-1/5}\left( \frac{n_0}{10^9\rm{cm}^{-3}} \right)^{3/10} \left( \frac{T}{10^7\rm{K}} \right)^{17/10} \rm{G} \label{eq:estb} \\
L_{\rm theor}=10^9\left( \frac{EM}{10^{48}\rm{cm}^{-3}} \right)^{3/5} \left(\frac{n_0}{10^9\rm{cm}^{-3}}\right)^{-2/5} \left(\frac{T}{10^7\rm{K}}\right)^{-8/5}\rm{cm} \label{eq:estl}
\end{eqnarray}
where $B_{\rm theor}$ is the coronal magnetic field strength, $L_{\rm theor}$ is the loop length, $T$ is the temperature and $EM$ is the emission measure, defined as $EM=n^2L^3$. It should be noted that $n$ is the electron density of the evaporated plasma, whereas $n_0$ is the pre-flare coronal electron density around the reconnection region.
It has been shown empirically that the pre-flare coronal electron density does not vary significantly between flares and that $n_0\sim10^9 \rm cm^{-3}$ \citep{2000PhDYashiro,2001ApJ...550L.113Y}. The coronal magnetic field ($B$) and loop length ($L$) of a flare can therefore be calculated as functions of $EM$ and $T$. 
We propose the following two examples as potential applications of this calculation method.

%太陽でも有効
Firstly, the scaling law is important in the study of solar flares. Although it is believed that the magnetic field plays a significant role in flares, the mechanism is not completely understood, partly because of difficulties in measuring coronal magnetic fields. The scaling law is therefore helpful because it can be used to estimate coronal magnetic fields. For example, Watanabe et al. (in prep.) applied it to the study of flares associated with visible emission, called white-light flares. They statistically compared white-light flares with `non-white-light flares', which emit no white light. Using this scaling law, they suggested that white-light flares tend to be associated with strong coronal magnetic fields.

%恒星の研究でも有効
Secondly, the scaling law could also be applied in stellar flare observations. Using the scaling law, the unobservable physical quantities $B$ and $L$ of stellar flares can be obtained from the observable physical quantities $EM$ and $T$, which are measurable with soft X-ray observation (e.g., \textit{ASCA}, \textit{Chandra}, and \textit{XMM-Newton}).  In the past decade, many stellar and protostellar flares have been observed (e.g., \cite{1991ApJ...378..725H,1996PASJ...48L..87K,1998ApJ...503..894T,2010ApJ...714L..98K}), and recently, many superflares on solar-type stars have also been detected \citep{2012Natur.485..478M,2013ApJS..209....5S}. The application of the scaling law to stellar flares has been useful in these cases. The basic structures of stellar atmospheres are considered to be very similar to that of solar atmosphere \citep{1989SoPh..121....3H,2002ARA&A..40..217G}, and the flare mechanism is thought to be the same \citep{1996ApJ...468L..37H}. 
Moreover, the estimated coronal magnetic field strength is roughly consistent with the observed values for the Sun and stars, 40-300 G \citep{1973SoPh...33..445R,1985ARA&A..23..169D,1996ApJ...456..840T,1997Natur.387...56G}.
 For these reasons, the scaling law can be applied to stellar flares and has in fact been used as a diagnostic tool (e.g., \cite{2004PASJ...56..813Y,2004A&A...419..653G,2005A&A...436.1041M,2016PASJ...68...90T}).

%では実証
To apply the scaling law of \authorcite{1999ApJ...526L..49S} (\yearcite{1999ApJ...526L..49S}, \yearcite{2002ApJ...577..422S}) to solar and stellar flares, it is necessary to test its validity. \citet{2002ApJ...579L..45Y} investigated the length scales of solar flares and arcades by using the scaling law (Equation ($\ref{eq:estl}$)), and they showed that the estimated values matched the observed values well. However, no one has tested its accuracy in calculating the coronal magnetic field strength (Equation ($\ref{eq:estb}$)) because of the difficulty involved in measuring this parameter observationally. \citet{2008ApJ...672..659A} commented on the necessity of statistical research on magnetic field strengths to test the scaling law. An effective test would involve investigating its validity on solar flares whose structures are spatially resolved.
In this paper, we describe analysis of 77 solar flares from the Hinode Flare Catalogue \citep{2012SoPh..279..317W}, shown in Table $\ref{table:1}$-$\ref{table:3}$. The flares of this catalogue occurred in between 2011 and 2014, and were observed with \textit{Hinode} satellite \citep{2007SoPh..243....3K}. For each flare, we present empirical estimation of the global coronal magnetic field strength of the reconnection region using \textit{Solar Dynamics Observatory} (\textit{SDO}) / Helioseismic and Magnetic Imager (HMI; \cite{2012SoPh..275..207S}) and Atmospheric Imaging Assembly (AIA; \cite{2012SoPh..275...17L}) data and discuss tests of the scaling law on the coronal magnetic field strengths and the length scales of the flares.  We demonstrate how to estimate these values theoretically and observationally in Section $\ref{sec:ana}$ and show the results of our tests of the scaling law in Section $\ref{sec:test}$.

%ほかのスケーリングもある。
We should note here that other methods of estimating the loop length of stellar flares have been proposed. For example, \citet{1991A&A...241..197S} derived a scaling law for estimating flare loop lengths from flare intensity decay time ($\tau_{\rm decay}$) based on the theory of flare loop cooling, $\tau_{\rm decay} \propto L/T^{1/2}$. \citet{1997A&A...325..782R} corrected this scaling law by incorporating the effect of the flare heating and also tested the validity of the scaling law using solar images. Hence, their estimation method has been widely used to estimate the stellar flare loop lengths \citep{1998A&A...334.1028R}. In contrast to the scaling law of \citet{1991A&A...241..197S}, the application of the \authorcite{1999ApJ...526L..49S} (\yearcite{1999ApJ...526L..49S}, \yearcite{2002ApJ...577..422S}) scaling law requires measurements of the observable quantities $T$ and $EM$ only at the flare peak. The detailed time evolutions of these quantities are not necessary in their model, which is advantageous since it is difficult to perform high-time-resolution observations of stellar flares. We compare the two scaling laws in Appendix \ref{sec:app}.

\section{Analysis}\label{sec:ana}
\subsection{Theoretical Estimation}\label{sec:est}
\begin{figure}
\begin{center}
\FigureFile(80mm,50mm){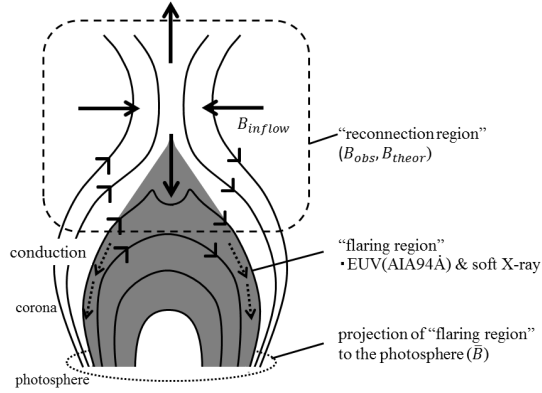}
\end{center}
\caption{Simple illustration of a magnetic reconnection model of flares. In this paper, reconnection regions are defined as global coronae in which magnetic energy is converted into kinetic and thermal energy, and flaring regions are defined as flaring coronal regions with bright soft X-ray or EUV emission. To estimate the coronal magnetic field strength of the reconnection region ($B_{\rm theor}$), we measured the photospheric magnetic field strength under the flaring region ($\overline{B}$; inside the dotted line). }
\label{fig:imgmap}
\end{figure}

Assuming that the coronal electron density does not vary significantly between flare events ($n_{0} \sim 10^9 \rm cm^{-3}$), the coronal magnetic field strength ($B_{\rm theor}$) and magnetic loop length ($L_{\rm theor}$) of reconnection regions can be theoretically estimated as functions of $T$ and $EM$, using Equations ($\ref{eq:estb}$) and ($\ref{eq:estl}$). As illustrated in Figure $\ref{fig:imgmap}$, reconnection regions are defined in this paper as global coronae in which magnetic energy is converted into kinetic and thermal energy, and flaring regions are defined as flaring coronal regions with bright soft X-ray or EUV emission. $T$ and $EM$ were calculated using a filter ratio method with \textit{GOES} soft X-ray fluxes. In this method, we used the program $goes\_chianti\_tem$ that is incorporated into the Solar SoftWare (SSW) package in IDL. We calculated the physical quantities required when $EM$ was at a maximum, as the $T$ value calculated by the filter ratio method is weighted by $EM$.

\subsection{Observational Estimation}\label{sec:obs}

\begin{figure}
\FigureFile(160mm,50mm){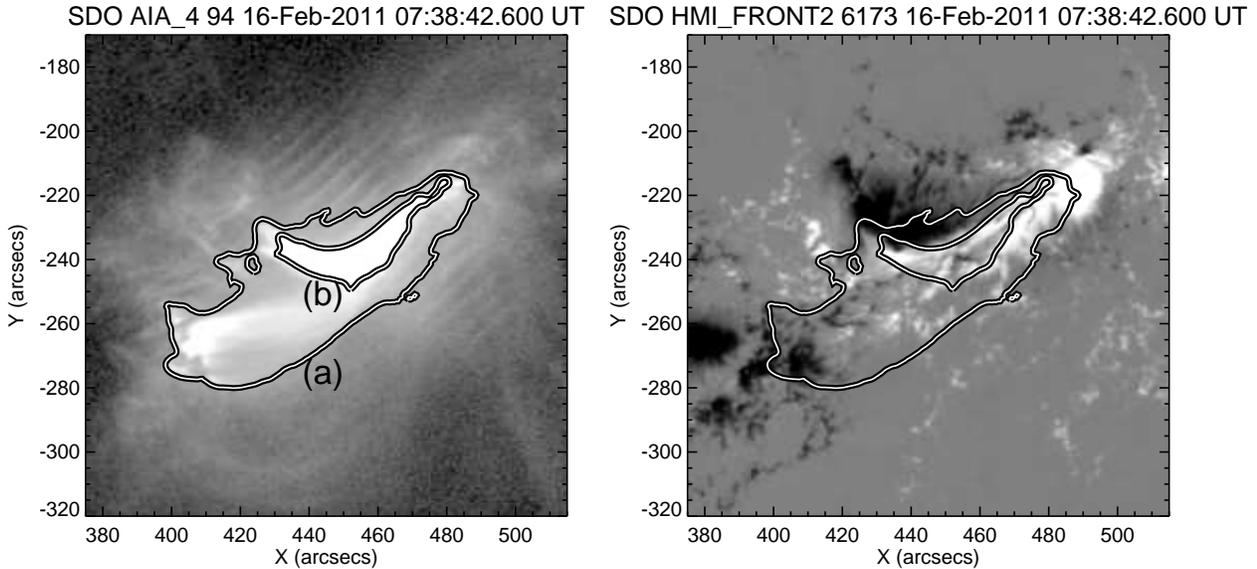}
\caption{Left: EUV image obtained with \textit{SDO}/AIA 94{\AA} on February 14, 2011. The image was observed at the peak time of \textit{GOES} soft X-ray, and we regarded flaring regions as regions with the intensities $>$30 $\rm DNs^{-1}$ (a; outer contour) and $>$500 $\rm DNs^{-1}$ (b; inner contour) in the image of AIA 94$\rm \AA$. Right: Image of the line-of-sight photospheric magnetic field obtained with \textit{SDO}/HMI Magnetgram at the same time as the left image. As illustrated in Figure $\ref{fig:imgmap}$, we measured the mean of the absolute value of magnetic field strength ($\overline{B}$) in the flaring region, and calculated the coronal magnetic field of the reconnection regions ($B_{\rm obs}$) using the empirical relation $B_{\rm obs}=\overline{B}/3$. }
\label{fig:map}
\end{figure}

Using an empirical relation, we observationally estimated the global coronal magnetic field strength where the magnetic energy was released. The analysis method was as follows. First, we defined the flaring regions in two different ways for comparison. The contours portrayed in the left of Figure $\ref{fig:map}$ show the flaring regions, defined as (a) regions with intensities $>$30 $\rm DNs^{-1}$ and (b) regions with intensities $>$500 $\rm DNs^{-1}$ in the images taken with \textit{SDO}/AIA 94{\AA} at the flare peak. Second, the mean of the absolute values of the photospheric magnetic field strength inside the projections of the flaring regions to the photosphere ($\overline{B}$) were given by \textit{SDO}/HMI Magnetogram (see, Figure $\ref{fig:imgmap}$ and the right panel of Figure \ref{fig:map}). The HMI images of very strong magnetic field regions could sometimes become saturated, so we replaced the saturated regions with the averages of the surrounding values. It is empirically known that the coronal magnetic field strengths of reconnection regions, $B_{\rm obs}$, are smaller by factors of $\sim$3 than photospheric ones, $\overline{B}$ (e.g., \cite{2005ApJ...632.1184I,2006ApJ...637.1122N}), and we therefore observationally estimated the coronal magnetic field as $B_{\rm obs}=\overline{B}/3$.

Although the coronal magnetic field strength can be extrapolated in the other ways (e.g., using a potential field or a force free field), we did not use these methods. 
It is not necessary to concern ourselves with measuring the coronal magnetic field strength exactly, since the scaling law itself still involves some error. In the analysis discussed in this report, we examined whether there was a positive correlation between the theoretically estimated magnetic field strength $B_{\rm theor}$ and the roughly measured one $B_{\rm obs}$. In addition, we defined the observed magnetic loop length scale $L_{\rm obs}$ as the square root of the area of the flaring region for simplicity and examined the relation between $L_{\rm theor}$ and $L_{\rm obs}$.

\section{Testing}\label{sec:test}
\subsection{Coronal Magnetic Field of the Flaring Region}\label{sec:testmag}

\begin{figure}
\begin{minipage}{0.5\hsize}
\begin{center}
\FigureFile(80mm,50mm){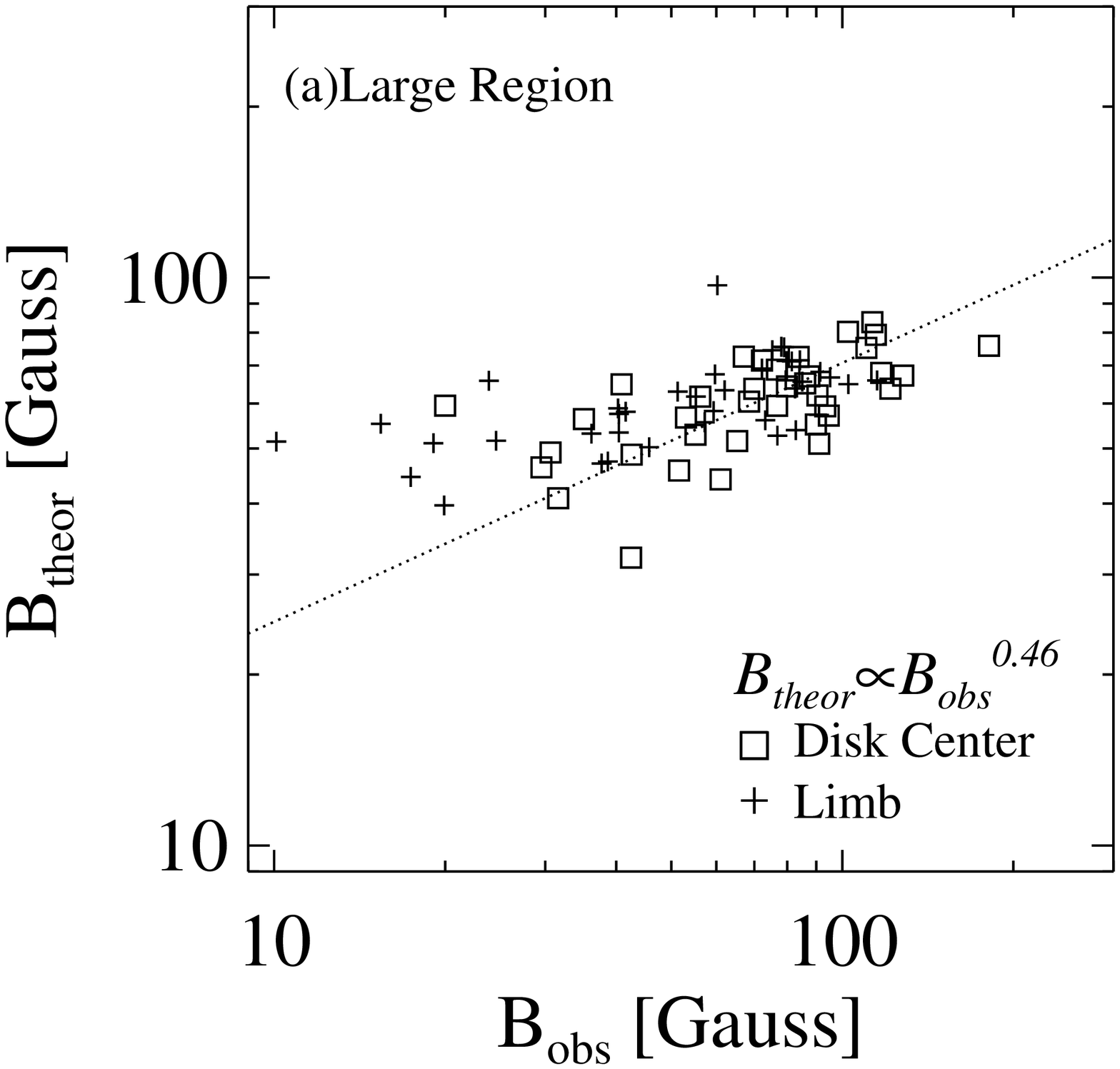}
\end{center}
\end{minipage}
\begin{minipage}{0.5\hsize}
\begin{center}
\FigureFile(80mm,50mm){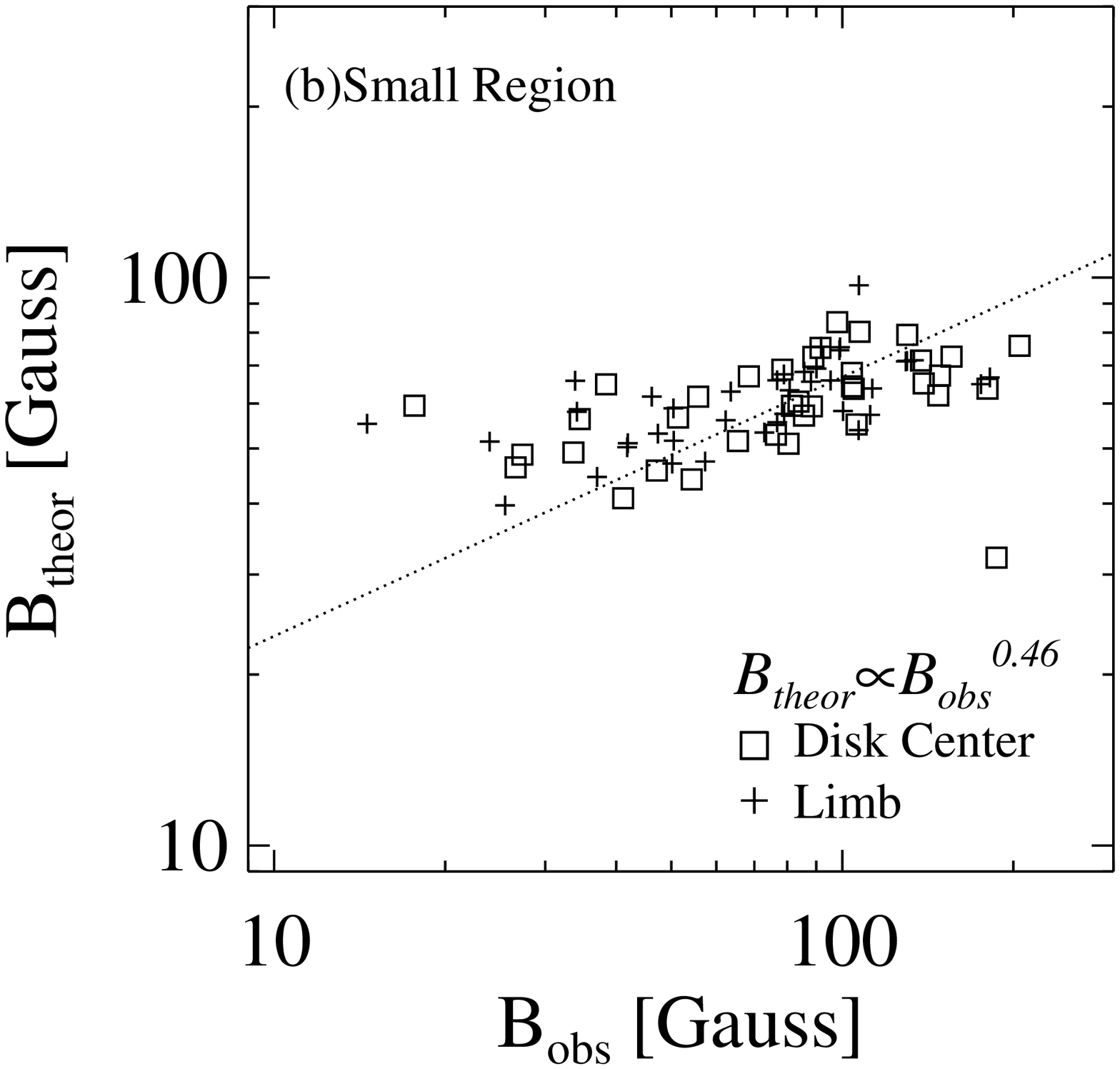}
\end{center}
\end{minipage}
\caption{Comparisons between the measured coronal magnetic field strength (based on \textit{SDO}/HMI) and the theoretically calculated one (based on the theory of \citet{2002ApJ...577..422S} and \textit{GOES} X-ray data). The left figure shows case (a), where flaring regions are large and are defined by a 30 $\rm DNs^{-1}$ contour, and the right is case (b), where flaring regions are small and are defined by 500 $\rm DNs^{-1}$.  Squares indicate flares with radial distances from the solar center of less than 600 arcsec, and crosses indicate those at distances greater than 600 arcsec. The dotted line is a fit for only the flares in the disk center, which was obtained using an ordinary least square bisector, and the power law index is 0.46 for both lines.}
\label{fig:bb}
\end{figure}

We compared the theoretically estimated coronal magnetic field strength calculated using the scaling law ($B_{\rm theor}$) with the observationally estimated one described in Section $\ref{sec:obs}$ ($B_{\rm obs}$). Figure $\ref{fig:bb}$ shows comparisons between $B_{\rm theor}$ and $B_{\rm obs}$. The left and right figures correspond to the case (a) and (b), respectively. The squares and crosses indicate the flares around the disk center and the solar limb, respectively. The disk center is defined as the area whose radial distance from the solar center is less than 600 arcsec (solar radius $\sim$1000 arcsec).
As expected, there is a positive correlation between $B_{\rm theor}$ and $B_{\rm obs}$, regardless of the definitions of the flaring regions.

We obtained this positive correlation using only simple analysis, which supports the validity of the initial assumption that flares are caused by magnetic reconnection, as well as supporting the application to stellar flares. The scaling law makes it possible to estimate the magnetic field strengths of unresolved stellar flares without high-resolution spectroscopy or high-time-resolution photometry, as mentioned in Section $\ref{sec:int}$. Such a simple method would be helpful for studying stellar flares, especially in large-sky surveys of nearby active flare stars. Statistical studies of stellar magnetic field strengths calculated using this scaling law may provide clues to understanding stellar activity.

As demonstrated in Figure $\ref{fig:bb}$, $B_{\rm theor}$ is not in proportional to $B_{\rm obs}$, although it should be proportional if the scaling law is correct. There are two possible reasons for this discrepancy. 
The first is related to the problem of the filter ratio method. The filter ratio method for \textit{GOES} incorporated into SSWIDL tends to estimate temperatures closer to 1 MK, which may have resulted in the narrow range of the theoretically estimated values as shown in Figure $\ref{fig:bb}$.
The second reason is the method of extrapolating $B_{\rm obs}$, which was done using the simple empirical relation $B_{\rm obs}=\overline{B}/3$. Moreover, the `photospheric' magnetic field strength was measured inside the regions defined by the `coronal' EUV brightness (\textit{SDO}/AIA 94{\AA}), making the estimated magnetic field strength smaller, especially for flares in the limb.
The flares in the limb and disk center are plotted individually in Figure $\ref{fig:bb}$, which shows that the observationally estimated values for flares in the limb tend to differ significantly from the theoretically estimated values. 
By an ordinary least-squares bisector fitting \citep{1990ApJ...364..104I}, the power law index was calculated to be $\sim$0.37 for all of the flares and $\sim$ 0.46 for the flares in the disk center in case (a), and $\sim$0.39 for all of the flares and $\sim$0.46 for the flares in the disk center in case (b).  

\subsection{Flare Loop Length}\label{sec:testlen}

\begin{figure}
\begin{minipage}{0.5\hsize}
\begin{center}
\FigureFile(80mm,50mm){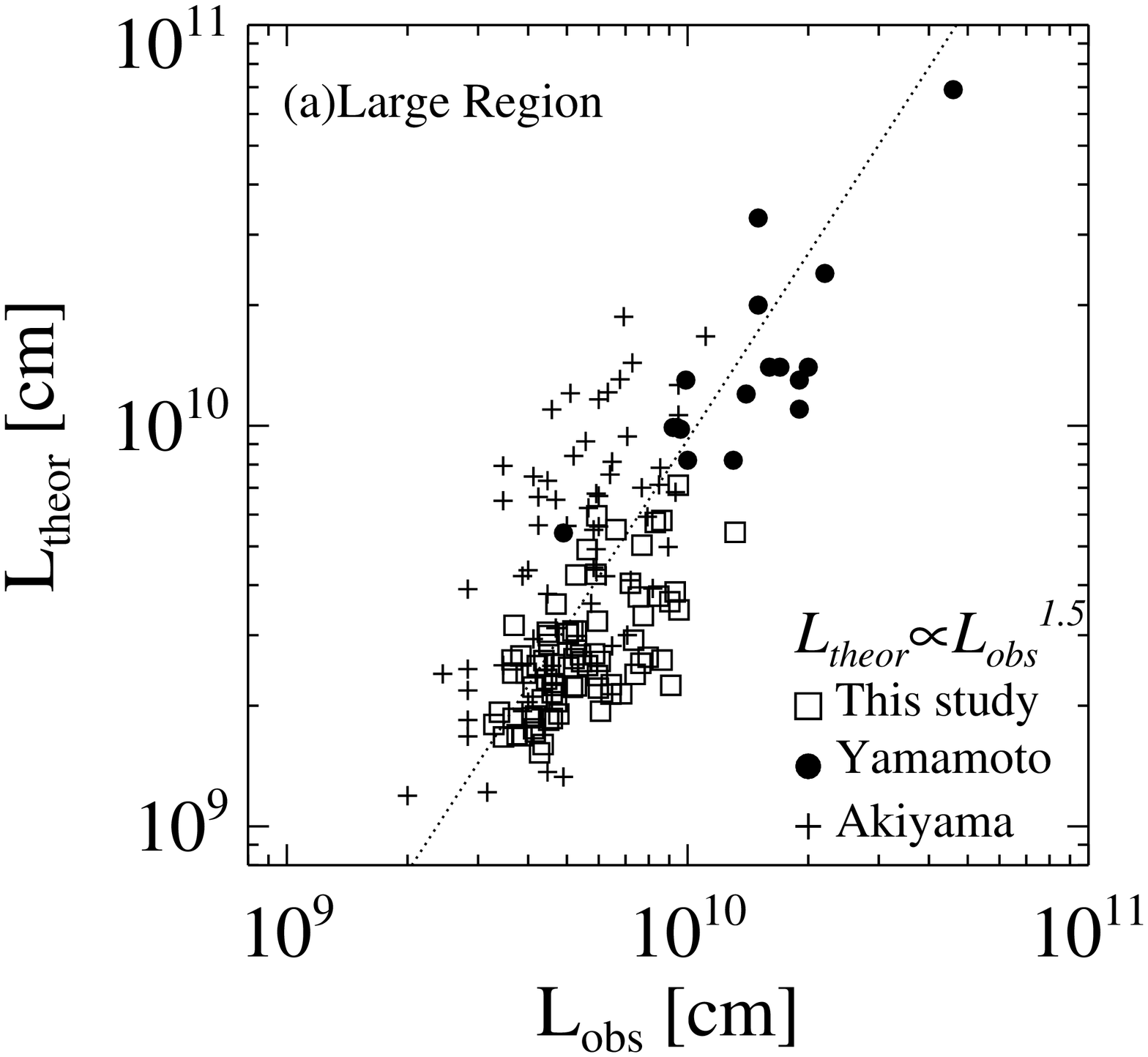}
\end{center}
\end{minipage}
\begin{minipage}{0.5\hsize}
\begin{center}
\FigureFile(80mm,50mm){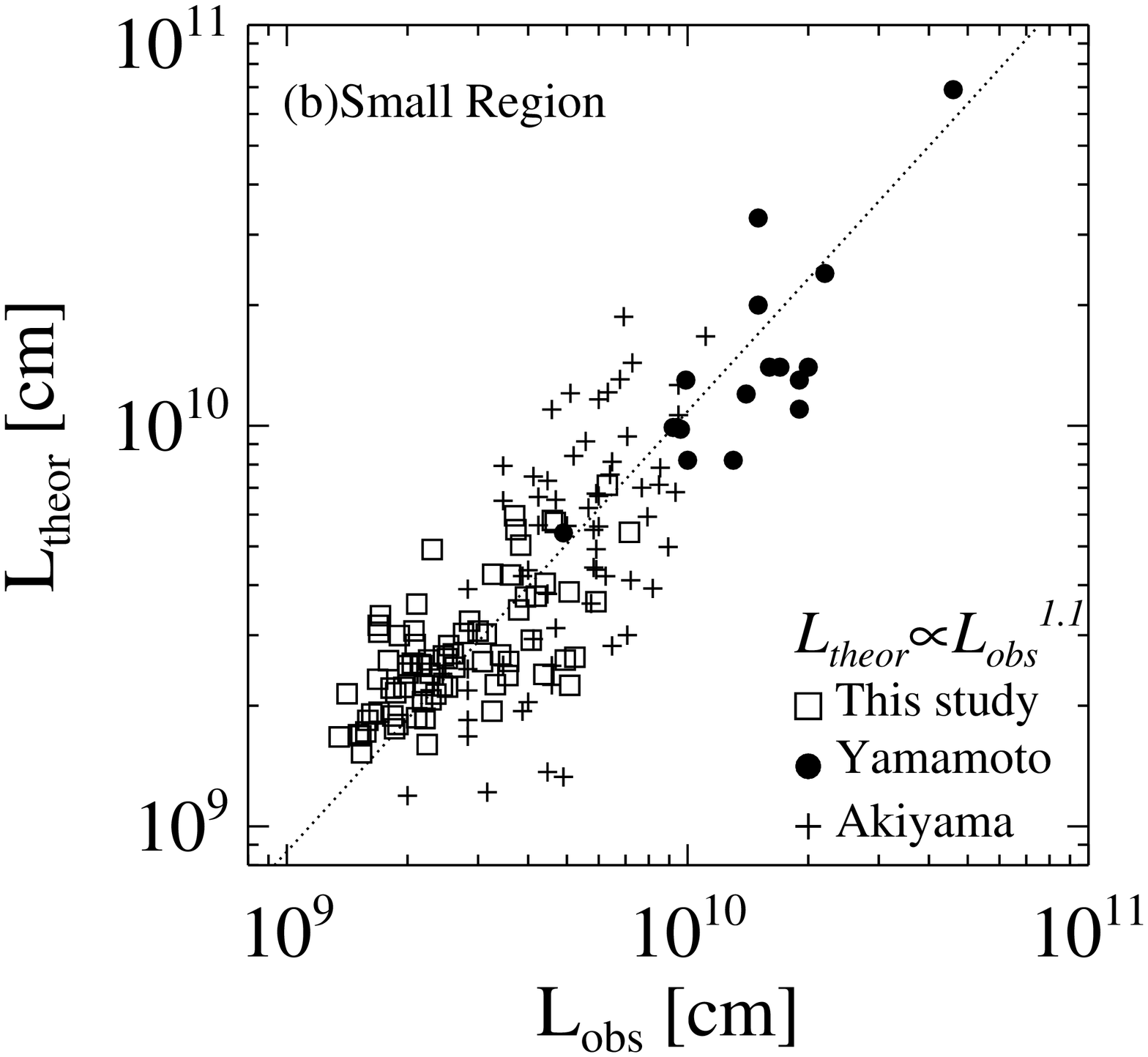}
\end{center}
\end{minipage}
\caption{Comparisons between the observationally estimated length scales ($L_{\rm obs}$) and the theoretically estimated ones ($L_{\rm theor}$) from \citet{2002ApJ...577..422S}. Squares correspond to flares listed in this paper, crosses indicate flares taken from \citet{2001PhDthesis}, and filled circles show arcades from \citet{2002ApJ...579L..45Y}. The left figure is case (a) and the right is case (b). $L_{\rm obs}$ matches $L_{\rm theor}$ well in case (b) and the power law index is $\sim$1.1. }
\label{fig:ll}
\end{figure}

\citet{2002ApJ...579L..45Y} previously tested the scaling law for C-, M- and X-class flares and arcades and reported a strong correlation between $L_{\rm theor}$ and $L_{\rm obs}$. They calculated $T$ and $EM$ using a filter ratio method with $Yohkoh$'s soft X-ray images, and the length scales of the flares and arcades were defined as the square root of the area of the flaring regions and the height of the arcades, respectively. Considering applications to the stellar flares, we examined the validity of the scaling law using X-ray fluxes measured with \textit{GOES} satellites, which is unable to resolve the structures.

As described in Section \ref{sec:obs}, the loop length scales of the flares, $L_{\rm obs}$, were measured as the square roots of the areas of the flaring regions with bright EUV emission in cases (a) and (b). 
Figure $\ref{fig:ll}$ shows comparisons between $L_{\rm obs}$ and $L_{\rm theor}$. 
The open squares correspond to the M- and X-class flares discussed in this paper. 
The crosses indicate the C-, M- and X-class flare data measured by \citet{2001PhDthesis} and the filled circles show the arcades measured by \citet{2002ApJ...579L..45Y}. 
In case (a), in which the measured area is defined by a lower threshold (30 $\rm DNs^{-1}$), the measured loop length scale $L_{\rm obs}$ tends to be larger than the theoretically estimated $L_{\rm theor}$. 
On the other hand, we can see that $L_{\rm theor}$ roughly matches $L_{\rm obs}$ with $\sigma$ $\sim$ 0.56, in case (b), where the measured area consists of bright flaring components. The power law index is $\sim$0.89 for our data and $\sim$1.10 for all of the data, including that of \citet{2001PhDthesis}  and \citet{2002ApJ...579L..45Y}.

These results show that the scaling law can be used to obtain accurate estimations of the  magnetic loop lengths of flares. 
In combination with photometry to estimate stellar spot sizes, this method would be helpful for constraining the magnetic field structures of unresolved stellar flares.

\section{Summary and Discussion}\label{sec:sum}
\authorcite{1999ApJ...526L..49S} (\yearcite{1999ApJ...526L..49S}, \yearcite{2002ApJ...577..422S}) suggested a method of estimating the coronal magnetic field strengths and magnetic loop length scales of solar and stellar flares from observations with an instrument that is unable to resolve their structure. In this paper, we described tests of the scaling law on spatially resolved solar flares with \textit{GOES} and \textit{SDO}/HMI and AIA data and found a positive correlation between the theoretically estimated physical quantities obtained with the scaling law and the observationally estimated values. We consider this analysis to support the theory of magnetic reconnection indirectly and to enable the application of the scaling law in the two scenarios described in Section \ref{sec:int}.

First, the scaling law provides an important suggestion regarding the unsolved mechanism of white-light flares by showing that the coronal magnetic fields of flares with white-light emission tend to be stronger than that of flares with no white light \citep{2016...Watanabe...prep}. This fact can be used to constrain white-light flare models, such as the model in which the main source of the flare optical continuum is the chromospheric condensation formed in the response of the chromosphere to flare heating \citep{1985ApJ...289..414F,1999ApJ...521..906A}. The reconnection with the strong magnetic field can lead to strong white-light emission.

Moreover, this analysis enables the application of the scaling law to the unresolved stellar flares and the determination of unobservable physical quantities. 
We should take into account the following two features regarding the application of the scaling law to stellar flares. Firstly, it is necessary to measure the maximum $T$ and $EM$ values of the flares as described in Section $\ref{sec:ana}$. In stellar flare observations, $T$ does not vary significantly during flares, regardless of the time cadence of the observations, while the maximum $EM$ may be reduced by a factor of 2-3 with long-exposure-time observations (e.g., \cite{2000ApJ...532.1089T}). This results in an increase in the estimated loop length by a factor of 1.5-1.9 (see, Equation ($\ref{eq:estl}$)) and does not affect the estimation. Secondly, it may be necessary to measure the pre-flare coronal density $n_0$, especially in the case of stellar flares, because $n_0$ may be higher by an order of magnitude for active stellar atmospheres and the estimated loop length can then be lower by a factor of $\sim$3.

Finally, we emphasize the usefulness of this method for stellar observations as follows: 
\begin{enumerate}
\item It can be used to estimate unobservable physical quantities theoretically, such as the coronal magnetic field strengths and loop lengths of unresolved stellar flares.
\item The application of the scaling law requires only photometric observations with an instrument that has a high sensitivity and moderate spectral resolution. Such a simple method would be useful in a large-sky survey of nearby active stars.
\item Additionally, the scaling law requires measurements of observed quantities only at the flare peak, while other scaling laws require time-evolution data of the rising or decay phases of flares. As we noted, the estimation for the scaling law presented does not require high-time-cadence observations, which is advantageous, especially in the case of impulsive flares.  
\end{enumerate}

\bigskip
Acknowledgement: We acknowledge with thanks Y. Notsu and S. Takasao for their contribution of fruitful comments on our work. We are grateful to the \textit{SDO}/AIA and HMI teams. \textit{SDO} is part of NASA’s Living with a Star Program. This work was partly carried out by using the Hinode Flare Catalogue ($http://st4a.stelab.nagoya-u.ac.jp/hinode\_flare/$), which is maintained by ISAS/JAXA and the Institute for Space-Earth Environmental Research, Nagoya University. \textit{Hinode} is a Japanese mission developed and launched by ISAS/JAXA with NAOJ as a domestic partner and NASA and STFC (UK) as international partners. This work was supported by JSPS KAKENHI Grant Numbers JP16H03955, JP15K17772, JP15K17622 and JP16H01187. The authors would like to thank Enago ($www.enago.jp$) for English language review.

\appendix
\section{Comparison with \citet{1997A&A...325..782R}}\label{sec:app}

We examined the validity of a scaling law proposed by \citet{1997A&A...325..782R} by applying it to the same flare catalogue used in this paper. As explained in Section 1, \citet{1991A&A...241..197S} first derived this scaling law for a flare loop in the decay phase. By incorporating the effect of heating during the decay phase into the scaling law, \citet{1997A&A...325..782R} corrected it as follows:
\begin{equation}
L=\frac{\tau_{LC}\sqrt{T}}{\alpha F(\zeta)}
\end{equation}
where $\tau_{LC}$ is a light curve decay time, $\alpha$ is $3.7\times 10^{-4} \rm cm^{-1} sK^{1/2}$, $\zeta$ is the slope in the $\sqrt{EM}$-$T$ diagram in the decay phase, and $F(\zeta)$ is the correction function:
\begin{equation}
F(\zeta)=\frac{c_a}{\zeta/\zeta_a-1}+q_a
\end{equation}
where $c_a=5.4 \pm 1.5$, $\zeta_a=0.25 \pm 0.04$, and $q_a=0.52\pm0.16$. In this study, we used \textit{GOES} soft X-ray fluxes to measure $T$, $\tau_{LC}$ and $\zeta$. As described by \citet{1997A&A...325..782R}, $\tau_{LC}$ is defined as the time taken for the \textit{GOES} count to decrease by $1/e$ from its maximum value. The slope $\zeta$ of the $\sqrt{EM}$-$T$ diagram was measured with a linear regression method, and the measured ranges were determined by eye as the ranges in which the slopes appeared straight in the diagrams. The measured values are listed in Table 1. We selected events that met the following condition:
\begin{enumerate}
\item They were listed in this paper and \citet{1997A&A...325..782R}.
\item They satisfied the conditions in which the \citet{1997A&A...325..782R} scaling law holds.
\item $\zeta \le 1.2$ (a range consistent with \citet{1997A&A...325..782R}).
\end{enumerate}
We then compared the estimation methods of \authorcite{1999ApJ...526L..49S} (\yearcite{1999ApJ...526L..49S}, \yearcite{2002ApJ...577..422S}) with that of \citet{1997A&A...325..782R}. Note that the $\zeta$ values in many flares were unmeasurable (see Table 1).

\begin{figure}
\begin{minipage}{0.5\hsize}
\begin{center}
\FigureFile(80mm,50mm){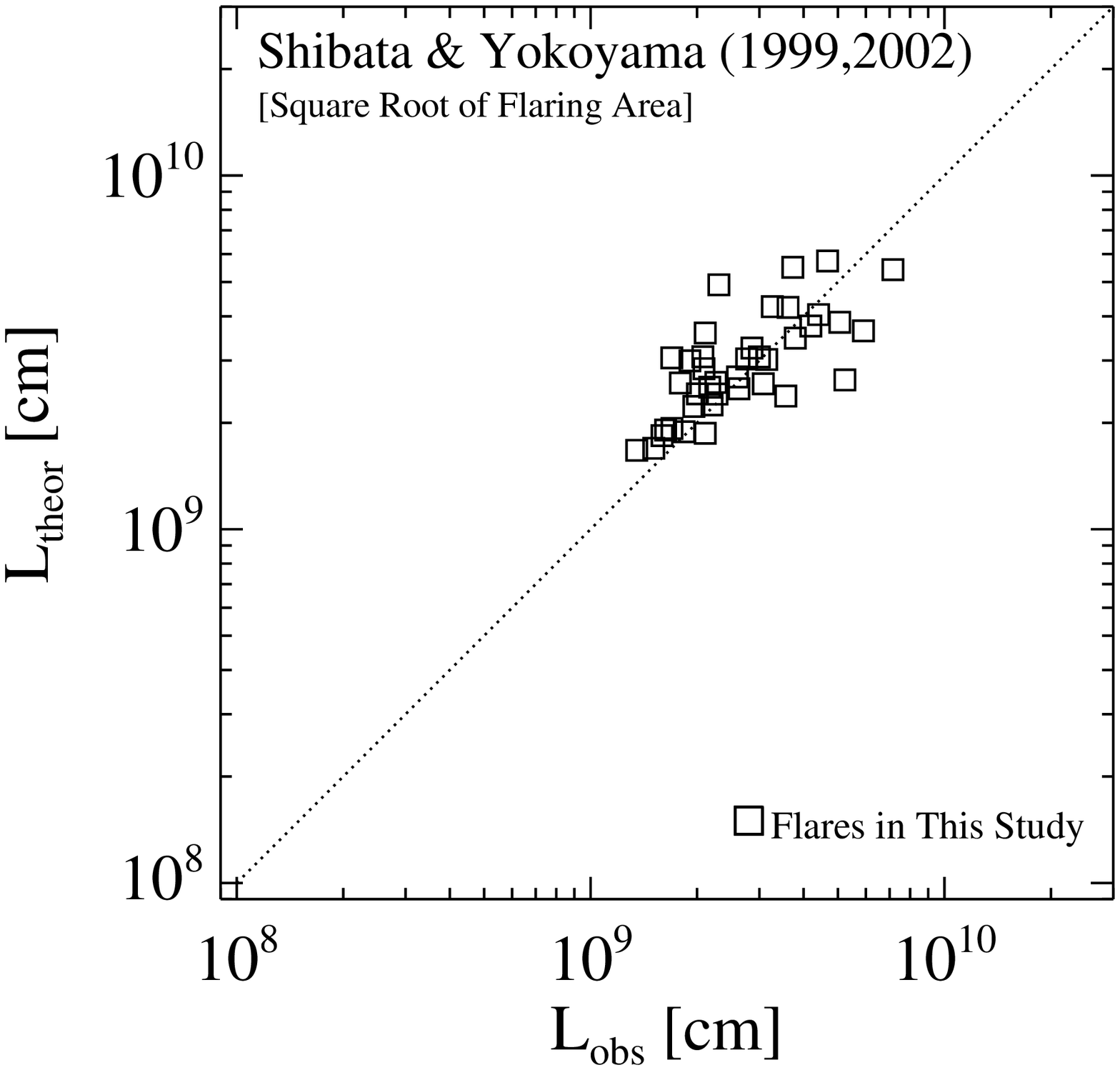}
\end{center}
\end{minipage}
\begin{minipage}{0.5\hsize}
\begin{center}
\FigureFile(80mm,50mm){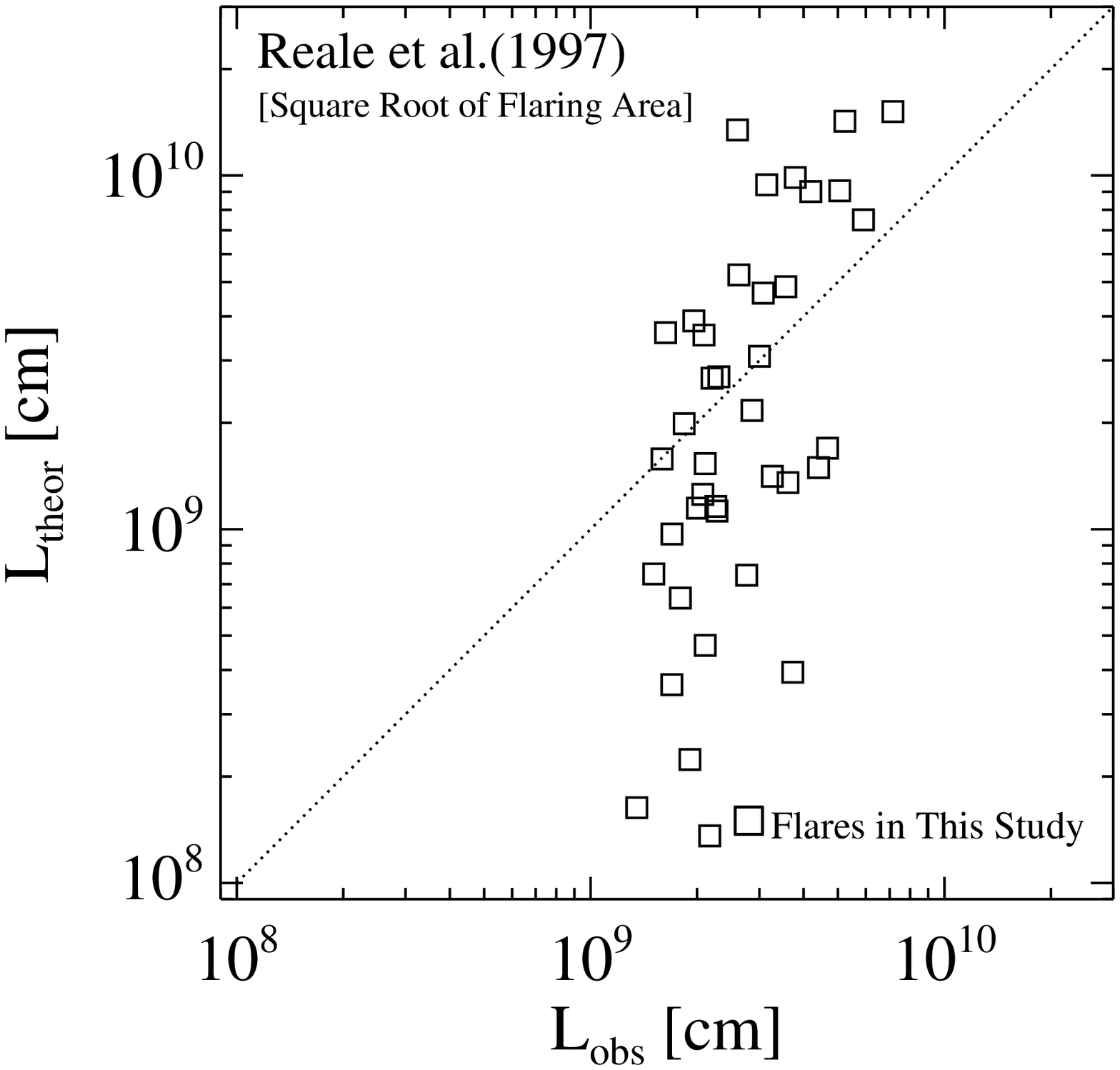}
\end{center}
\end{minipage}
\caption{Left and right figures show comparisons between the observationally measured and theoretically estimated loop lengths using the scaling law of \citet{2002ApJ...577..422S} and \citet{1997A&A...325..782R}, respectively. The loop lengths were measured as the square roots of the flaring regions defined by 500 $\rm DNs^{-1}$ (b; inner contour) of the brightness from AIA 94$\rm \AA$ as described in Section $\ref{sec:ana}$.}
\label{fig:fig4}
\end{figure}

\begin{figure}
\begin{minipage}{0.5\hsize}
\begin{center}
\FigureFile(80mm,50mm){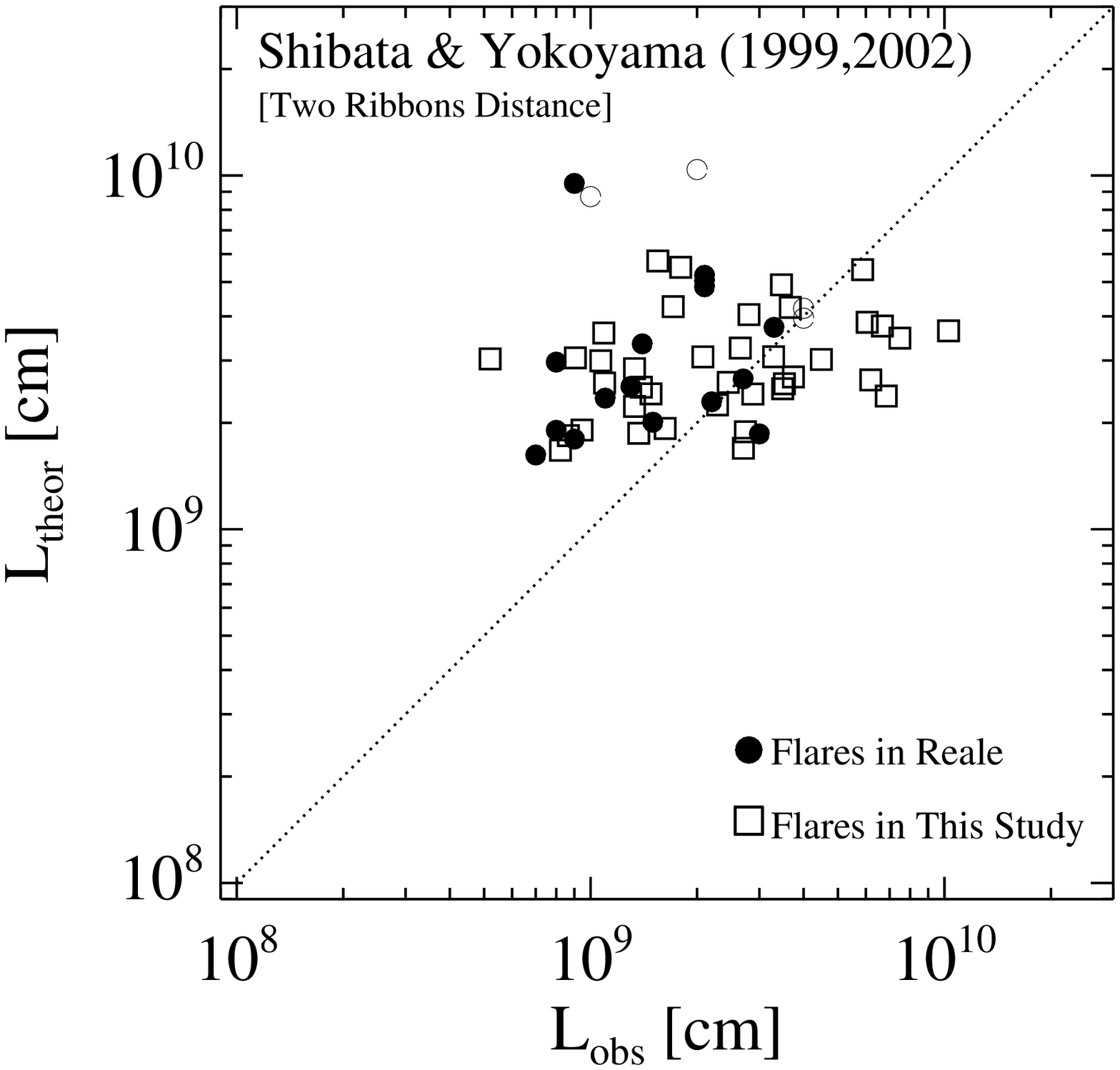}
\end{center}
\end{minipage}
\begin{minipage}{0.5\hsize}
\begin{center}
\FigureFile(80mm,50mm){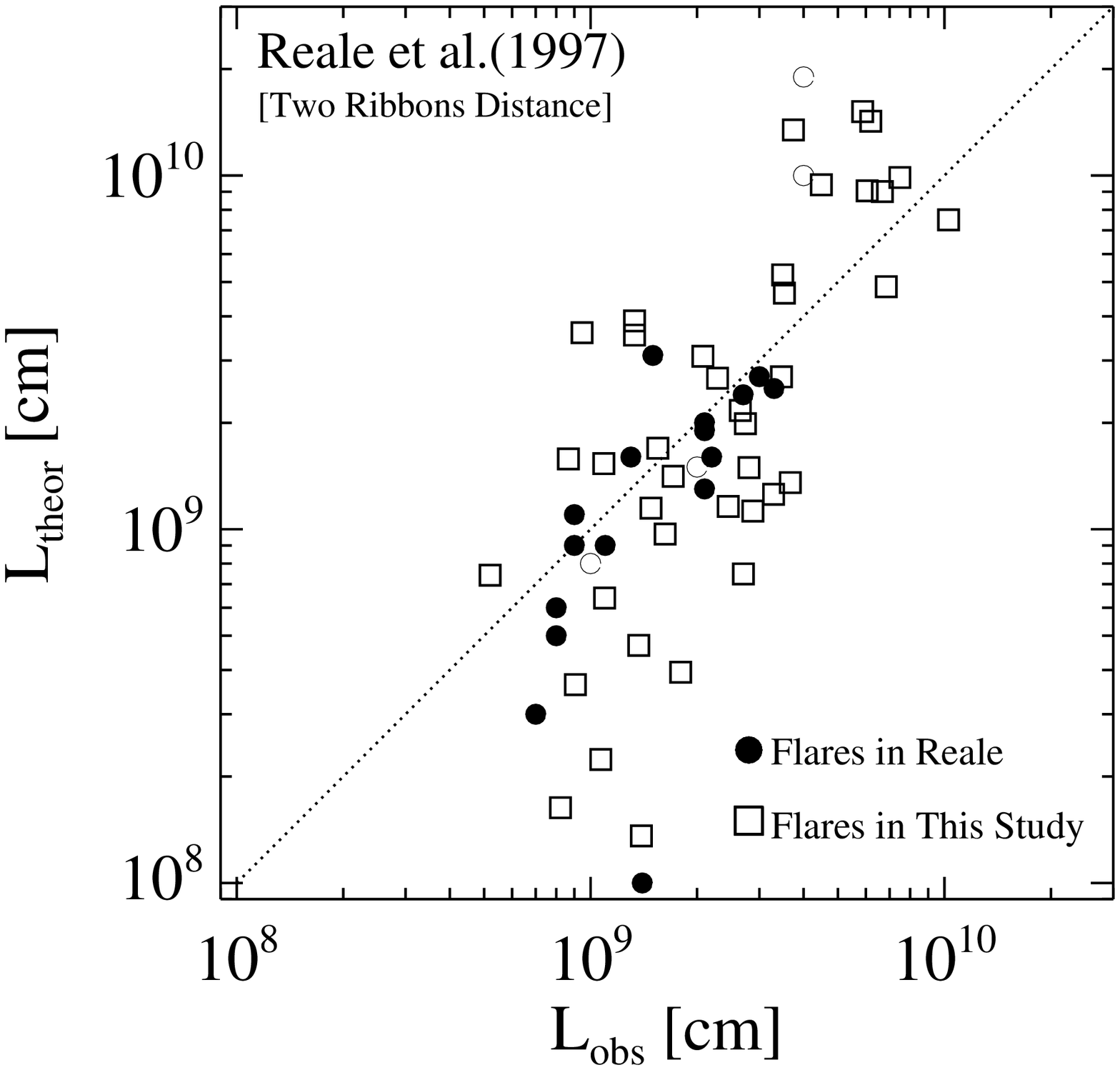}
\end{center}
\end{minipage}
\caption{Left and right figures show comparisons between the observationally measured  and theoretically estimated loop lengths using the scaling law of \citet{2002ApJ...577..422S} and \citet{1997A&A...325..782R}, respectively. The loop lengths were measured as the distance of two ribbons in images taken by AIA 1600$\rm \AA$. Since it was difficult to measure the distances between the two ribbons at the flare peaks due to the saturation of the images, we analyzed the images 5 min after the flare peaks. The distances were measured by eye between the brightest points of the flare ribbons.}
\label{fig:fig5}
\end{figure}

Figures \ref{fig:fig4} and \ref{fig:fig5} show comparisons of the two scaling laws. 
In this analysis, we measured the flaring loop lengths in two ways. 
The loop length in Figure \ref{fig:fig4} is defined as the square root of the area of the flaring regions measured with \textit{SDO}/AIA 94{\AA}, and the loop length in Figure \ref{fig:fig5} is the distance between two ribbons measured with \textit{SDO}/AIA 1600{\AA}. 
The left and right panels in each figure show the results of  \authorcite{1999ApJ...526L..49S} (\yearcite{1999ApJ...526L..49S}, \yearcite{2002ApJ...577..422S}) and \citet{1997A&A...325..782R}. 
The unfilled squares correspond to the flares listed in this paper, and the circles indicate those selected by \citet{1997A&A...325..782R} to test their scaling law. 
The filled circles are events for which the flare loops were clearly measured, and the unfilled ones are events for which flare loops were too complex to measure. 
The estimated values obtained with both scaling laws are consistent with the true values, except for in some individual events. 
In the right panels of Figure \ref{fig:fig4} and \ref{fig:fig5}, the loop length estimates from \citet{1997A&A...325..782R} are scattered. 
This scattering is present because, while \citet{1997A&A...325..782R} measured $\zeta$ using the spatially resolved images taken with Yohkoh, it is difficult to measure well-defined $\zeta$ values with  \textit{GOES} satellite, which observes the entire solar surface. 
The overall solar emission measure $EM_{s}\sim10^{48-49}\rm cm^{-3}$ is comparable to the flare value of $EM\sim10^{49}\rm cm^{-3}$, which makes it difficult to measure $\zeta$ precisely, especially in observations of flares on solar-type stars (G stars).
Another reason is that there is a difference between the instruments used by \citet{1997A&A...325..782R} and those used by this study. \citet{1997A&A...325..782R} used the light curves with \textit{Yohkoh} Al 11.4 \rm{$\mu m$} filter band, which is sensitive to a few mega-Kelvin, while we used \textit{GOES} 1-8{\AA} pass band, which is sensitive to a few tens of mega-Kelvin.

Finally, we compared the practicality of the two scaling laws. \citet{1997A&A...325..782R}'s scaling law requires time-evolution data for the flares to estimate physical quantities, while that of \authorcite{1999ApJ...526L..49S} (\yearcite{1999ApJ...526L..49S}, \yearcite{2002ApJ...577..422S}) does not require high-time-cadence observations, as noted in Section $\ref{sec:sum}$. This is a significant advantage because it is difficult to observe stellar flares with high time cadence. Moreover, methods that use decay or rising phases introduce additional uncertainties, since the soft X-ray light curves of flares are superpositions of light curves of several flaring loops that reconnect one after another. 
In terms of application to stellar observations, we consider the scaling law of \citet{2002ApJ...577..422S} to be more useful for estimating the physical parameters of unresolved stellar flares.

\setcounter{figure}{0}
\setcounter{table}{0}

\begin{landscape}
\begin{table*}
\caption{Physical paremeters of flares.}
\begin{center}
\scriptsize
\begin{tabular}{lccccccccccccc}
\hline
\footnotesize{peak time} & \footnotesize{\textit{GOES}} & \footnotesize{Loc.}\commenta & \footnotesize{$T_6$}\commentb & \footnotesize{$EM_{48}$}\commentc & \footnotesize{$B_{\rm theor}$}\commentd & \footnotesize{$B_{\rm obs}(a)$}\commente & \footnotesize{$B_{\rm obs}(b)$}\commentf & \footnotesize{$L_{\rm theor}$}\commentd & \footnotesize{$L_{\rm obs}(a)$}\commente & \footnotesize{$L_{\rm obs}(b)$}\commentf & \footnotesize{$L_{\rm Reale}$}\commentg & $\tau_{\rm SXR}$\commenth & $\zeta$\commenti\\
 & \footnotesize{class} & \footnotesize{[deg]}  & \footnotesize{[K]} & \footnotesize{[$\rm cm^{-3}$] } & \footnotesize{[G]} & \footnotesize{[G]} & \footnotesize{[G]} & \footnotesize{$\rm [10^9cm]$} & \footnotesize{$\rm [10^9cm]$} & \footnotesize{$\rm [10^9cm]$} & \footnotesize{$\rm [10^9cm]$} & \footnotesize{[sec]} & \\
\hline
\footnotesize{2011/02/14-17:26} & \footnotesize{M2.2} & \footnotesize{S20W04} & \footnotesize{15.5} & \footnotesize{1.2} & \footnotesize{63.7} & \footnotesize{70.0} & \footnotesize{104.7} & \footnotesize{2.2} & \footnotesize{5.2} & \footnotesize{2.0} & \footnotesize{3.9} & \footnotesize{770} & \footnotesize{1.04} \\
\footnotesize{2011/02/15-01:56} & \footnotesize{X2.2} & \footnotesize{S20W10} & \footnotesize{19.6} & \footnotesize{10.3} & \footnotesize{62.0} & \footnotesize{90.4} & \footnotesize{147.5} & \footnotesize{5.5} & \footnotesize{6.6} & \footnotesize{3.7} & \footnotesize{0.4} & \footnotesize{747} & \footnotesize{0.30} \\
\footnotesize{2011/02/16-07:44} & \footnotesize{M1.1} & \footnotesize{S19W29} & \footnotesize{14.3} & \footnotesize{0.6} & \footnotesize{63.7} & \footnotesize{121.5} & \footnotesize{180.1} & \footnotesize{1.7} & \footnotesize{3.9} & \footnotesize{1.5} & \footnotesize{$-$} & \footnotesize{$-$} & \footnotesize{$-$} \\
\footnotesize{2011/02/18-10:11} & \footnotesize{M6.6} & \footnotesize{S21W55} & \footnotesize{18.6} & \footnotesize{3.4} & \footnotesize{71.2} & \footnotesize{81.5} & \footnotesize{129.4} & \footnotesize{3.1} & \footnotesize{4.5} & \footnotesize{1.7} & \footnotesize{0.4} & \footnotesize{233} & \footnotesize{0.42} \\
\footnotesize{2011/02/18-13:03} & \footnotesize{M1.4} & \footnotesize{S21W55} & \footnotesize{15.7} & \footnotesize{0.8} & \footnotesize{71.3} & \footnotesize{79.7} & \footnotesize{129.9} & \footnotesize{1.7} & \footnotesize{3.5} & \footnotesize{1.4} & \footnotesize{0.2} & \footnotesize{274} & \footnotesize{0.32} \\
\footnotesize{2011/09/23-22:15} & \footnotesize{M1.6} & \footnotesize{N12E56} & \footnotesize{12.7} & \footnotesize{1.0} & \footnotesize{47.4} & \footnotesize{38.7} & \footnotesize{57.3} & \footnotesize{2.7} & \footnotesize{5.9} & \footnotesize{2.6} & \footnotesize{13.5} & \footnotesize{3209} & \footnotesize{0.97} \\
\footnotesize{2011/11/02-22:01} & \footnotesize{M4.3} & \footnotesize{N20E77} & \footnotesize{17.2} & \footnotesize{2.1} & \footnotesize{68.2} & \footnotesize{91.5} & \footnotesize{85.7} & \footnotesize{2.6} & \footnotesize{5.2} & \footnotesize{2.5} & \footnotesize{$-$} & \footnotesize{1384} & \footnotesize{$-$} \\
\footnotesize{2011/11/03-11:11} & \footnotesize{M2.5} & \footnotesize{N20E70} & \footnotesize{15.9} & \footnotesize{1.3} & \footnotesize{65.6} & \footnotesize{85.0} & \footnotesize{88.1} & \footnotesize{2.2} & \footnotesize{4.7} & \footnotesize{2.2} & \footnotesize{2.7} & \footnotesize{917} & \footnotesize{0.64} \\
\footnotesize{2011/11/03-20:27} & \footnotesize{X1.9} & \footnotesize{N21E64} & \footnotesize{21.5} & \footnotesize{8.7} & \footnotesize{75.5} & \footnotesize{78.1} & \footnotesize{91.8} & \footnotesize{4.3} & \footnotesize{5.9} & \footnotesize{3.3} & \footnotesize{1.4} & \footnotesize{407} & \footnotesize{0.66} \\
\footnotesize{2011/11/05-03:31} & \footnotesize{M3.7} & \footnotesize{N20E47} & \footnotesize{15.8} & \footnotesize{1.5} & \footnotesize{63.8} & \footnotesize{82.6} & \footnotesize{112.9} & \footnotesize{2.4} & \footnotesize{7.4} & \footnotesize{4.4} & \footnotesize{$-$} & \footnotesize{2209} & \footnotesize{$-$} \\
\footnotesize{2011/11/05-20:38} & \footnotesize{M1.8} & \footnotesize{N21E37} & \footnotesize{17.0} & \footnotesize{0.9} & \footnotesize{79.3} & \footnotesize{114.5} & \footnotesize{130.1} & \footnotesize{1.6} & \footnotesize{4.4} & \footnotesize{2.2} & \footnotesize{$-$} & \footnotesize{1427} & \footnotesize{$-$} \\
\footnotesize{2011/12/31-13:15} & \footnotesize{M2.4} & \footnotesize{S25E46} & \footnotesize{17.0} & \footnotesize{1.2} & \footnotesize{74.5} & \footnotesize{75.3} & \footnotesize{98.8} & \footnotesize{1.9} & \footnotesize{3.4} & \footnotesize{1.7} & \footnotesize{1.0} & \footnotesize{352} & \footnotesize{0.61} \\
\footnotesize{2011/12/31-16:26} & \footnotesize{M1.5} & \footnotesize{S25E42} & \footnotesize{15.1} & \footnotesize{0.9} & \footnotesize{65.9} & \footnotesize{79.5} & \footnotesize{76.8} & \footnotesize{1.9} & \footnotesize{3.7} & \footnotesize{2.2} & \footnotesize{$-$} & \footnotesize{636} & \footnotesize{$-$} \\
\footnotesize{2012/01/17-04:53} & \footnotesize{M1.0} & \footnotesize{N18E53} & \footnotesize{14.0} & \footnotesize{0.6} & \footnotesize{61.7} & \footnotesize{55.2} & \footnotesize{46.2} & \footnotesize{1.7} & \footnotesize{4.2} & \footnotesize{1.6} & \footnotesize{$-$} & \footnotesize{1429} & \footnotesize{$-$} \\
\footnotesize{2012/01/18-19:12} & \footnotesize{M1.7} & \footnotesize{N17E32} & \footnotesize{14.4} & \footnotesize{1.0} & \footnotesize{59.4} & \footnotesize{76.8} & \footnotesize{81.5} & \footnotesize{2.2} & \footnotesize{4.6} & \footnotesize{1.9} & \footnotesize{$-$} & \footnotesize{1476} & \footnotesize{$-$} \\
\footnotesize{2012/01/27-18:37} & \footnotesize{X1.7} & \footnotesize{N33W85} & \footnotesize{15.9} & \footnotesize{9.0} & \footnotesize{44.6} & \footnotesize{17.4} & \footnotesize{37.0} & \footnotesize{7.1} & \footnotesize{9.5} & \footnotesize{6.3} & \footnotesize{$-$} & \footnotesize{1480} & \footnotesize{$-$} \\
\footnotesize{2012/03/05-04:05} & \footnotesize{X1.1} & \footnotesize{N19E58} & \footnotesize{18.2} & \footnotesize{5.0} & \footnotesize{63.0} & \footnotesize{51.3} & \footnotesize{63.6} & \footnotesize{4.0} & \footnotesize{7.2} & \footnotesize{4.4} & \footnotesize{1.5} & \footnotesize{2785} & \footnotesize{0.31} \\
\footnotesize{2012/03/06-12:41} & \footnotesize{M2.1} & \footnotesize{N21E40} & \footnotesize{16.0} & \footnotesize{1.1} & \footnotesize{67.9} & \footnotesize{117.1} & \footnotesize{104.0} & \footnotesize{2.0} & \footnotesize{4.7} & \footnotesize{2.2} & \footnotesize{$-$} & \footnotesize{1105} & \footnotesize{$-$} \\
\footnotesize{2012/03/06-21:11} & \footnotesize{M1.3} & \footnotesize{N16E30} & \footnotesize{16.1} & \footnotesize{0.7} & \footnotesize{75.8} & \footnotesize{181.2} & \footnotesize{205.1} & \footnotesize{1.5} & \footnotesize{4.3} & \footnotesize{1.5} & \footnotesize{$-$} & \footnotesize{$-$} & \footnotesize{$-$} \\
\footnotesize{2012/05/07-14:31} & \footnotesize{M1.9} & \footnotesize{S20W49} & \footnotesize{14.0} & \footnotesize{1.0} & \footnotesize{55.3} & \footnotesize{15.4} & \footnotesize{14.6} & \footnotesize{2.4} & \footnotesize{6.0} & \footnotesize{3.6} & \footnotesize{4.8} & \footnotesize{1777} & \footnotesize{0.66} \\
\footnotesize{2012/05/09-12:32} & \footnotesize{M4.7} & \footnotesize{N13E31} & \footnotesize{17.8} & \footnotesize{2.3} & \footnotesize{71.4} & \footnotesize{72.2} & \footnotesize{137.5} & \footnotesize{2.6} & \footnotesize{4.4} & \footnotesize{1.8} & \footnotesize{0.6} & \footnotesize{260} & \footnotesize{0.56} \\
\footnotesize{2012/05/09-14:08} & \footnotesize{M1.8} & \footnotesize{N06E22} & \footnotesize{13.4} & \footnotesize{1.1} & \footnotesize{51.0} & \footnotesize{91.1} & \footnotesize{80.4} & \footnotesize{2.6} & \footnotesize{3.6} & \footnotesize{2.3} & \footnotesize{1.2} & \footnotesize{374} & \footnotesize{0.68} \\
\footnotesize{2012/05/09-21:05} & \footnotesize{M4.1} & \footnotesize{N12E26} & \footnotesize{17.7} & \footnotesize{2.0} & \footnotesize{72.5} & \footnotesize{83.8} & \footnotesize{88.9} & \footnotesize{2.4} & \footnotesize{3.7} & \footnotesize{2.0} & \footnotesize{1.1} & \footnotesize{346} & \footnotesize{0.66} \\
\footnotesize{2012/05/10-20:26} & \footnotesize{M1.7} & \footnotesize{N12E12} & \footnotesize{15.7} & \footnotesize{0.9} & \footnotesize{68.8} & \footnotesize{76.7} & \footnotesize{78.4} & \footnotesize{1.9} & \footnotesize{4.1} & \footnotesize{2.1} & \footnotesize{0.5} & \footnotesize{616} & \footnotesize{0.35} \\
\footnotesize{2012/06/13-13:17} & \footnotesize{M1.2} & \footnotesize{S16E18} & \footnotesize{13.5} & \footnotesize{0.7} & \footnotesize{56.7} & \footnotesize{53.1} & \footnotesize{51.4} & \footnotesize{1.9} & \footnotesize{6.1} & \footnotesize{3.2} & \footnotesize{$-$} & \footnotesize{$-$} & \footnotesize{$-$} \\
\footnotesize{2012/07/05-03:36} & \footnotesize{M4.7} & \footnotesize{S18W29} & \footnotesize{18.0} & \footnotesize{2.3} & \footnotesize{72.6} & \footnotesize{67.1} & \footnotesize{155.7} & \footnotesize{2.6} & \footnotesize{7.6} & \footnotesize{2.1} & \footnotesize{$-$} & \footnotesize{249} & \footnotesize{$-$} \\
\footnotesize{2012/07/05-21:45} & \footnotesize{M1.6} & \footnotesize{S12W46} & \footnotesize{13.4} & \footnotesize{1.0} & \footnotesize{51.6} & \footnotesize{24.6} & \footnotesize{50.5} & \footnotesize{2.5} & \footnotesize{5.6} & \footnotesize{2.0} & \footnotesize{$-$} & \footnotesize{$-$} & \footnotesize{$-$} \\
\footnotesize{2012/07/06-01:40} & \footnotesize{M2.9} & \footnotesize{S18W41} & \footnotesize{16.3} & \footnotesize{1.5} & \footnotesize{66.7} & \footnotesize{95.2} & \footnotesize{181.8} & \footnotesize{2.3} & \footnotesize{4.5} & \footnotesize{1.7} & \footnotesize{$-$} & \footnotesize{176} & \footnotesize{$-$} \\
\footnotesize{2012/07/06-08:23} & \footnotesize{M1.5} & \footnotesize{S17W40} & \footnotesize{15.0} & \footnotesize{0.8} & \footnotesize{64.9} & \footnotesize{102.5} & \footnotesize{175.4} & \footnotesize{1.9} & \footnotesize{4.2} & \footnotesize{1.8} & \footnotesize{2.0} & \footnotesize{727} & \footnotesize{0.64} \\
\footnotesize{2012/07/06-23:08} & \footnotesize{X1.1} & \footnotesize{S13W59} & \footnotesize{16.6} & \footnotesize{5.5} & \footnotesize{53.1} & \footnotesize{36.2} & \footnotesize{47.4} & \footnotesize{4.9} & \footnotesize{5.6} & \footnotesize{2.3} & \footnotesize{2.7} & \footnotesize{542} & \footnotesize{0.86} \\
\footnotesize{2012/07/08-12:09} & \footnotesize{M1.4} & \footnotesize{S21W69} & \footnotesize{15.8} & \footnotesize{0.8} & \footnotesize{71.5} & \footnotesize{84.2} & \footnotesize{133.6} & \footnotesize{1.7} & \footnotesize{3.8} & \footnotesize{1.5} & \footnotesize{0.7} & \footnotesize{274} & \footnotesize{0.62} \\
\footnotesize{2012/07/12-16:49} & \footnotesize{X1.4} & \footnotesize{S13W03} & \footnotesize{20.5} & \footnotesize{3.4} & \footnotesize{83.5} & \footnotesize{112.9} & \footnotesize{98.0} & \footnotesize{2.6} & \footnotesize{8.0} & \footnotesize{5.2} & \footnotesize{14.2} & \footnotesize{3536} & \footnotesize{0.77} \\
\footnotesize{2012/07/14-04:58} & \footnotesize{M1.0} & \footnotesize{S22W36} & \footnotesize{9.7} & \footnotesize{0.7} & \footnotesize{32.1} & \footnotesize{42.5} & \footnotesize{186.8} & \footnotesize{3.4} & \footnotesize{7.8} & \footnotesize{1.7} & \footnotesize{$-$} & \footnotesize{$-$} & \footnotesize{$-$} \\
\footnotesize{2012/10/23-03:17} & \footnotesize{X1.8} & \footnotesize{S13E58} & \footnotesize{24.2} & \footnotesize{6.7} & \footnotesize{96.9} & \footnotesize{60.3} & \footnotesize{106.9} & \footnotesize{3.0} & \footnotesize{5.0} & \footnotesize{2.8} & \footnotesize{0.7} & \footnotesize{303} & \footnotesize{0.51} \\
\footnotesize{2013/05/02-05:10} & \footnotesize{M1.1} & \footnotesize{N10W26} & \footnotesize{12.4} & \footnotesize{0.7} & \footnotesize{49.2} & \footnotesize{30.6} & \footnotesize{33.6} & \footnotesize{2.2} & \footnotesize{4.1} & \footnotesize{2.5} & \footnotesize{$-$} & \footnotesize{651} & \footnotesize{$-$} \\
\footnotesize{2013/05/03-16:55} & \footnotesize{M1.3} & \footnotesize{N10W38} & \footnotesize{12.3} & \footnotesize{0.8} & \footnotesize{46.3} & \footnotesize{29.5} & \footnotesize{26.6} & \footnotesize{2.6} & \footnotesize{6.1} & \footnotesize{3.6} & \footnotesize{$-$} & \footnotesize{$-$} & \footnotesize{$-$} \\
\footnotesize{2013/05/15-01:40} & \footnotesize{X1.2} & \footnotesize{N12E64} & \footnotesize{15.8} & \footnotesize{6.6} & \footnotesize{47.1} & \footnotesize{37.7} & \footnotesize{50.2} & \footnotesize{6.0} & \footnotesize{5.9} & \footnotesize{3.7} & \footnotesize{$-$} & \footnotesize{909} & \footnotesize{$-$} \\
\hline
\end{tabular}
\end{center}
\label{table:1}
\end{table*}
\end{landscape}

\begin{landscape}
\begin{table*}
\caption{Physical paremeter of flares (continued).}
\begin{center}
\scriptsize
\begin{tabular}{lccccccccccccc}
\hline
\footnotesize{peak time} & \footnotesize{\textit{GOES}} & \footnotesize{Loc.}\commenta & \footnotesize{$T_6$}\commentb & \footnotesize{$EM_{48}$}\commentc & \footnotesize{$B_{\rm theor}$}\commentd & \footnotesize{$B_{\rm obs}(a)$}\commente & \footnotesize{$B_{\rm obs}(b)$}\commentf & \footnotesize{$L_{\rm theor}$}\commentd & \footnotesize{$L_{\rm obs}(a)$}\commente & \footnotesize{$L_{\rm obs}(b)$}\commentf & \footnotesize{$L_{\rm Reale}$}\commentg & $\tau_{\rm SXR}$\commenth & $\zeta$\commenti\\
 & \footnotesize{class} & \footnotesize{[deg]}  & \footnotesize{[K]} & \footnotesize{[$\rm cm^{-3}$] } & \footnotesize{[G]} & \footnotesize{[G]} & \footnotesize{[G]} & \footnotesize{$\rm [10^9cm]$} & \footnotesize{$\rm [10^9cm]$} & \footnotesize{$\rm [10^9cm]$} & \footnotesize{$\rm [10^9cm]$} & \footnotesize{[sec]} & \\
\hline
\footnotesize{2013/06/07-22:49} & \footnotesize{M5.9} & \footnotesize{S32W89} & \footnotesize{17.4} & \footnotesize{2.9} & \footnotesize{65.8} & \footnotesize{23.9} & \footnotesize{33.9} & \footnotesize{3.1} & \footnotesize{5.3} & \footnotesize{3.0} & \footnotesize{3.1} & \footnotesize{1204} & \footnotesize{0.58} \\
\footnotesize{2013/08/17-18:24} & \footnotesize{M3.3} & \footnotesize{S07W30} & \footnotesize{16.4} & \footnotesize{1.7} & \footnotesize{65.1} & \footnotesize{86.0} & \footnotesize{139.1} & \footnotesize{2.5} & \footnotesize{4.7} & \footnotesize{2.2} & \footnotesize{$-$} & \footnotesize{$-$} & \footnotesize{$-$} \\
\footnotesize{2013/10/22-00:22} & \footnotesize{M1.0} & \footnotesize{N06E17} & \footnotesize{13.6} & \footnotesize{0.6} & \footnotesize{59.3} & \footnotesize{93.3} & \footnotesize{88.4} & \footnotesize{1.8} & \footnotesize{4.1} & \footnotesize{1.9} & \footnotesize{$-$} & \footnotesize{571} & \footnotesize{$-$} \\
\footnotesize{2013/10/26-19:31} & \footnotesize{M3.1} & \footnotesize{S09E81} & \footnotesize{14.2} & \footnotesize{1.7} & \footnotesize{51.1} & \footnotesize{19.1} & \footnotesize{41.9} & \footnotesize{3.2} & \footnotesize{3.7} & \footnotesize{1.7} & \footnotesize{$-$} & \footnotesize{380} & \footnotesize{$-$} \\
\footnotesize{2013/10/28-15:01} & \footnotesize{M2.7} & \footnotesize{S08E28} & \footnotesize{15.3} & \footnotesize{1.5} & \footnotesize{60.5} & \footnotesize{68.5} & \footnotesize{83.7} & \footnotesize{2.5} & \footnotesize{4.2} & \footnotesize{2.2} & \footnotesize{0.1} & \footnotesize{141} & \footnotesize{0.37} \\
\footnotesize{2013/10/28-15:15} & \footnotesize{M4.4} & \footnotesize{S06E28} & \footnotesize{13.6} & \footnotesize{2.5} & \footnotesize{44.1} & \footnotesize{61.1} & \footnotesize{54.3} & \footnotesize{4.2} & \footnotesize{5.3} & \footnotesize{3.6} & \footnotesize{1.4} & \footnotesize{1030} & \footnotesize{0.42} \\
\footnotesize{2013/12/07-07:29} & \footnotesize{M1.2} & \footnotesize{S16W49} & \footnotesize{11.3} & \footnotesize{0.9} & \footnotesize{39.7} & \footnotesize{19.9} & \footnotesize{25.5} & \footnotesize{3.0} & \footnotesize{5.0} & \footnotesize{3.1} & \footnotesize{9.4} & \footnotesize{2289} & \footnotesize{0.97} \\
\footnotesize{2013/12/22-08:11} & \footnotesize{M1.9} & \footnotesize{S20W49} & \footnotesize{15.9} & \footnotesize{1.0} & \footnotesize{69.2} & \footnotesize{72.2} & \footnotesize{90.1} & \footnotesize{1.9} & \footnotesize{4.8} & \footnotesize{1.6} & \footnotesize{3.6} & \footnotesize{722} & \footnotesize{0.99} \\
\footnotesize{2013/12/22-15:12} & \footnotesize{M3.3} & \footnotesize{S19W56} & \footnotesize{15.4} & \footnotesize{1.8} & \footnotesize{58.2} & \footnotesize{59.4} & \footnotesize{100.3} & \footnotesize{2.8} & \footnotesize{4.5} & \footnotesize{2.1} & \footnotesize{3.5} & \footnotesize{739} & \footnotesize{1.02} \\
\footnotesize{2013/12/23-09:06} & \footnotesize{M1.6} & \footnotesize{S17W63} & \footnotesize{15.4} & \footnotesize{0.9} & \footnotesize{67.5} & \footnotesize{59.7} & \footnotesize{78.9} & \footnotesize{1.8} & \footnotesize{4.5} & \footnotesize{1.6} & \footnotesize{1.6} & \footnotesize{286} & \footnotesize{1.18} \\
\footnotesize{2013/12/31-21:58} & \footnotesize{M6.4} & \footnotesize{S16W35} & \footnotesize{17.4} & \footnotesize{3.1} & \footnotesize{64.3} & \footnotesize{79.9} & \footnotesize{104.0} & \footnotesize{3.3} & \footnotesize{6.0} & \footnotesize{2.9} & \footnotesize{2.2} & \footnotesize{1593} & \footnotesize{0.42} \\
\footnotesize{2014/01/01-18:52} & \footnotesize{M9.9} & \footnotesize{S14W47} & \footnotesize{15.9} & \footnotesize{5.1} & \footnotesize{50.3} & \footnotesize{45.7} & \footnotesize{41.8} & \footnotesize{5.0} & \footnotesize{7.7} & \footnotesize{3.8} & \footnotesize{$-$} & \footnotesize{1007} & \footnotesize{$-$} \\
\footnotesize{2014/01/04-10:25} & \footnotesize{M1.3} & \footnotesize{S05E48} & \footnotesize{13.6} & \footnotesize{0.8} & \footnotesize{56.1} & \footnotesize{73.2} & \footnotesize{62.3} & \footnotesize{2.1} & \footnotesize{4.3} & \footnotesize{2.3} & \footnotesize{$-$} & \footnotesize{$-$} & \footnotesize{$-$} \\
\footnotesize{2014/01/04-19:46} & \footnotesize{M4.0} & \footnotesize{S11E34} & \footnotesize{15.2} & \footnotesize{1.5} & \footnotesize{59.5} & \footnotesize{20.0} & \footnotesize{17.7} & \footnotesize{2.6} & \footnotesize{8.6} & \footnotesize{5.0} & \footnotesize{$-$} & \footnotesize{3219} & \footnotesize{$-$} \\
\footnotesize{2014/02/14-02:57} & \footnotesize{M2.3} & \footnotesize{S12W25} & \footnotesize{14.6} & \footnotesize{1.3} & \footnotesize{57.2} & \footnotesize{94.7} & \footnotesize{85.6} & \footnotesize{2.5} & \footnotesize{5.0} & \footnotesize{2.6} & \footnotesize{5.2} & \footnotesize{1198} & \footnotesize{0.94} \\
\footnotesize{2014/02/14-12:40} & \footnotesize{M1.6} & \footnotesize{S15W36} & \footnotesize{13.9} & \footnotesize{0.9} & \footnotesize{56.3} & \footnotesize{35.1} & \footnotesize{34.5} & \footnotesize{2.2} & \footnotesize{6.0} & \footnotesize{1.8} & \footnotesize{$-$} & \footnotesize{415} & \footnotesize{$-$} \\
\footnotesize{2014/03/13-19:19} & \footnotesize{M1.2} & \footnotesize{N15W87} & \footnotesize{12.9} & \footnotesize{0.8} & \footnotesize{51.4} & \footnotesize{10.1} & \footnotesize{24.0} & \footnotesize{2.3} & \footnotesize{6.5} & \footnotesize{3.3} & \footnotesize{$-$} & \footnotesize{1105} & \footnotesize{$-$} \\
\footnotesize{2014/06/12-22:16} & \footnotesize{M3.1} & \footnotesize{S20W55} & \footnotesize{14.2} & \footnotesize{1.0} & \footnotesize{57.5} & \footnotesize{40.5} & \footnotesize{79.0} & \footnotesize{2.2} & \footnotesize{9.1} & \footnotesize{5.1} & \footnotesize{$-$} & \footnotesize{3536} & \footnotesize{$-$} \\
\footnotesize{2014/06/16-00:01} & \footnotesize{M1.0} & \footnotesize{S19E08} & \footnotesize{11.1} & \footnotesize{0.7} & \footnotesize{40.9} & \footnotesize{31.6} & \footnotesize{41.1} & \footnotesize{2.7} & \footnotesize{3.8} & \footnotesize{2.5} & \footnotesize{$-$} & \footnotesize{2019} & \footnotesize{$-$} \\
\footnotesize{2014/08/01-18:03} & \footnotesize{M1.5} & \footnotesize{S10E11} & \footnotesize{12.3} & \footnotesize{0.9} & \footnotesize{45.7} & \footnotesize{51.6} & \footnotesize{47.1} & \footnotesize{2.7} & \footnotesize{5.4} & \footnotesize{3.4} & \footnotesize{$-$} & \footnotesize{3260} & \footnotesize{$-$} \\
\footnotesize{2014/10/21-13:38} & \footnotesize{M1.2} & \footnotesize{S14E35} & \footnotesize{14.8} & \footnotesize{0.8} & \footnotesize{64.9} & \footnotesize{40.9} & \footnotesize{38.4} & \footnotesize{1.8} & \footnotesize{3.3} & \footnotesize{1.9} & \footnotesize{$-$} & \footnotesize{129} & \footnotesize{$-$} \\
\footnotesize{2014/10/22-01:59} & \footnotesize{M8.7} & \footnotesize{S13E21} & \footnotesize{18.4} & \footnotesize{4.0} & \footnotesize{67.2} & \footnotesize{128.0} & \footnotesize{148.6} & \footnotesize{3.5} & \footnotesize{9.5} & \footnotesize{3.8} & \footnotesize{9.9} & \footnotesize{2297} & \footnotesize{0.84} \\
\footnotesize{2014/10/22-14:06} & \footnotesize{X1.6} & \footnotesize{S14E13} & \footnotesize{20.5} & \footnotesize{5.9} & \footnotesize{75.2} & \footnotesize{110.3} & \footnotesize{91.5} & \footnotesize{3.6} & \footnotesize{9.0} & \footnotesize{5.9} & \footnotesize{7.5} & \footnotesize{1710} & \footnotesize{0.77} \\
\footnotesize{2014/10/24-21:15} & \footnotesize{X3.1} & \footnotesize{S16W21} & \footnotesize{19.4} & \footnotesize{9.8} & \footnotesize{61.7} & \footnotesize{56.2} & \footnotesize{55.8} & \footnotesize{5.4} & \footnotesize{13.1} & \footnotesize{7.2} & \footnotesize{15.2} & \footnotesize{2646} & \footnotesize{1.00} \\
\footnotesize{2014/10/25-17:08} & \footnotesize{X1.0} & \footnotesize{S10W22} & \footnotesize{18.7} & \footnotesize{4.8} & \footnotesize{67.0} & \footnotesize{87.5} & \footnotesize{68.5} & \footnotesize{3.7} & \footnotesize{7.5} & \footnotesize{3.9} & \footnotesize{$-$} & \footnotesize{$-$} & \footnotesize{$-$} \\
\footnotesize{2014/10/26-10:56} & \footnotesize{X2.0} & \footnotesize{S18W40} & \footnotesize{21.9} & \footnotesize{7.3} & \footnotesize{80.2} & \footnotesize{102.3} & \footnotesize{107.2} & \footnotesize{3.8} & \footnotesize{8.5} & \footnotesize{4.2} & \footnotesize{9.0} & \footnotesize{1683} & \footnotesize{0.93} \\
\footnotesize{2014/10/26-20:21} & \footnotesize{M2.4} & \footnotesize{S15W45} & \footnotesize{15.7} & \footnotesize{1.2} & \footnotesize{65.9} & \footnotesize{115.1} & \footnotesize{95.4} & \footnotesize{2.1} & \footnotesize{6.4} & \footnotesize{2.4} & \footnotesize{$-$} & \footnotesize{$-$} & \footnotesize{$-$} \\
\footnotesize{2014/10/27-14:23} & \footnotesize{X2.0} & \footnotesize{S17W52} & \footnotesize{20.9} & \footnotesize{6.7} & \footnotesize{75.4} & \footnotesize{79.0} & \footnotesize{99.1} & \footnotesize{3.8} & \footnotesize{9.3} & \footnotesize{5.1} & \footnotesize{9.1} & \footnotesize{1763} & \footnotesize{0.92} \\
\footnotesize{2014/10/28-02:42} & \footnotesize{M3.4} & \footnotesize{S14W61} & \footnotesize{15.3} & \footnotesize{1.8} & \footnotesize{57.3} & \footnotesize{91.0} & \footnotesize{111.9} & \footnotesize{2.9} & \footnotesize{7.3} & \footnotesize{4.1} & \footnotesize{$-$} & \footnotesize{$-$} & \footnotesize{$-$} \\
\footnotesize{2014/10/29-10:01} & \footnotesize{M1.2} & \footnotesize{S18W77} & \footnotesize{13.0} & \footnotesize{0.7} & \footnotesize{52.7} & \footnotesize{76.9} & \footnotesize{$-$} & \footnotesize{2.1} & \footnotesize{6.8} & \footnotesize{1.4} & \footnotesize{$-$} & \footnotesize{$-$} & \footnotesize{$-$} \\
\footnotesize{2014/11/05-09:47} & \footnotesize{M7.9} & \footnotesize{N20E68} & \footnotesize{17.6} & \footnotesize{3.8} & \footnotesize{63.3} & \footnotesize{62.1} & \footnotesize{80.8} & \footnotesize{3.6} & \footnotesize{4.7} & \footnotesize{2.1} & \footnotesize{1.5} & \footnotesize{679} & \footnotesize{0.50} \\
\footnotesize{2014/11/05-19:44} & \footnotesize{M2.9} & \footnotesize{N17E65} & \footnotesize{15.1} & \footnotesize{1.4} & \footnotesize{58.9} & \footnotesize{40.3} & \footnotesize{50.4} & \footnotesize{2.6} & \footnotesize{5.4} & \footnotesize{3.1} & \footnotesize{4.7} & \footnotesize{2525} & \footnotesize{0.50} \\
\footnotesize{2014/11/07-02:44} & \footnotesize{M2.7} & \footnotesize{N17E50} & \footnotesize{14.3} & \footnotesize{1.5} & \footnotesize{53.9} & \footnotesize{82.8} & \footnotesize{106.8} & \footnotesize{2.8} & \footnotesize{5.3} & \footnotesize{2.5} & \footnotesize{$-$} & \footnotesize{$-$} & \footnotesize{$-$} \\
\footnotesize{2014/11/07-10:20} & \footnotesize{M1.0} & \footnotesize{N15E43} & \footnotesize{13.1} & \footnotesize{0.6} & \footnotesize{55.6} & \footnotesize{57.3} & \footnotesize{76.8} & \footnotesize{1.9} & \footnotesize{4.6} & \footnotesize{2.2} & \footnotesize{$-$} & \footnotesize{954} & \footnotesize{$-$} \\
\footnotesize{2014/11/07-17:26} & \footnotesize{X1.6} & \footnotesize{N14E36} & \footnotesize{17.3} & \footnotesize{7.9} & \footnotesize{52.9} & \footnotesize{55.1} & \footnotesize{76.4} & \footnotesize{5.7} & \footnotesize{8.3} & \footnotesize{4.7} & \footnotesize{1.7} & \footnotesize{798} & \footnotesize{0.52} \\
\footnotesize{2014/11/15-12:03} & \footnotesize{M3.2} & \footnotesize{S09E63} & \footnotesize{14.6} & \footnotesize{1.8} & \footnotesize{53.4} & \footnotesize{40.4} & \footnotesize{72.9} & \footnotesize{3.1} & \footnotesize{5.2} & \footnotesize{2.1} & \footnotesize{1.3} & \footnotesize{559} & \footnotesize{0.53} \\
\footnotesize{2014/11/15-20:46} & \footnotesize{M3.7} & \footnotesize{S13E63} & \footnotesize{15.5} & \footnotesize{2.0} & \footnotesize{58.0} & \footnotesize{41.6} & \footnotesize{34.1} & \footnotesize{3.0} & \footnotesize{4.5} & \footnotesize{1.9} & \footnotesize{0.2} & \footnotesize{276} & \footnotesize{0.35} \\
\hline
\end{tabular}
\end{center}
\label{table:2}
\end{table*}
\end{landscape}

\begin{landscape}
\begin{table*}
\caption{Physical paremeter of flares (continued).}
\begin{center}
\scriptsize
\begin{tabular}{lccccccccccccc}
\hline
\footnotesize{peak time} & \footnotesize{\textit{GOES}} & \footnotesize{Loc.}\commenta & \footnotesize{$T_6$}\commentb & \footnotesize{$EM_{48}$}\commentc & \footnotesize{$B_{\rm theor}$}\commentd & \footnotesize{$B_{\rm obs}(a)$}\commente & \footnotesize{$B_{\rm obs}(b)$}\commentf & \footnotesize{$L_{\rm theor}$}\commentd & \footnotesize{$L_{\rm obs}(a)$}\commente & \footnotesize{$L_{\rm obs}(b)$}\commentf & \footnotesize{$L_{\rm Reale}$}\commentg & $\tau_{\rm SXR}$\commenth & $\zeta$\commenti\\
 & \footnotesize{class} & \footnotesize{[deg]}  & \footnotesize{[K]} & \footnotesize{[$\rm cm^{-3}$] } & \footnotesize{[G]} & \footnotesize{[G]} & \footnotesize{[G]} & \footnotesize{$\rm [10^9cm]$} & \footnotesize{$\rm [10^9cm]$} & \footnotesize{$\rm [10^9cm]$} & \footnotesize{$\rm [10^9cm]$} & \footnotesize{[sec]} & \\
\hline
\footnotesize{2014/12/04-08:03} & \footnotesize{M1.3} & \footnotesize{S24W27} & \footnotesize{12.9} & \footnotesize{0.8} & \footnotesize{51.5} & \footnotesize{65.3} & \footnotesize{65.5} & \footnotesize{2.2} & \footnotesize{5.3} & \footnotesize{2.4} & \footnotesize{$-$} & \footnotesize{$-$} & \footnotesize{$-$} \\
\footnotesize{2014/12/19-09:44} & \footnotesize{M1.3} & \footnotesize{S19W27} & \footnotesize{12.6} & \footnotesize{0.8} & \footnotesize{48.8} & \footnotesize{42.6} & \footnotesize{27.3} & \footnotesize{2.4} & \footnotesize{4.0} & \footnotesize{2.3} & \footnotesize{1.1} & \footnotesize{1144} & \footnotesize{0.39} \\
\footnotesize{2014/12/20-00:28} & \footnotesize{X1.8} & \footnotesize{S21W24} & \footnotesize{18.0} & \footnotesize{8.9} & \footnotesize{55.1} & \footnotesize{89.9} & \footnotesize{105.9} & \footnotesize{5.8} & \footnotesize{8.6} & \footnotesize{4.6} & \footnotesize{$-$} & \footnotesize{2975} & \footnotesize{$-$} \\
\hline
\multicolumn{5}{l}{\commenta Locations where flares occurred.} \\
\multicolumn{5}{l}{\commentb Temperature in unit $10^6\rm K$ when emission measure is maximum value.} \\
\multicolumn{5}{l}{\commentc Emission measure in unit $10^{48}\rm cm^{-3}$.} \\
\multicolumn{5}{l}{\commentd Theoretically estimated physical quantities by \citet{2002ApJ...577..422S}.} \\
\multicolumn{5}{l}{\commente Observationally measured physical quantities in the case of (a).} \\
\multicolumn{5}{l}{\commentf Observationally measured physical quantities in the case of (b).} \\
\multicolumn{5}{l}{\commentg Theoretically estimated loop length by Reale et al. (1997).} \\
\multicolumn{5}{l}{\commenth Flare duration defined as e-folding decay time of soft X-ray intensity.} \\
\multicolumn{5}{l}{\commenti The trajectory of $\sqrt{EM}-T$ diagram.} \\
\end{tabular}
\end{center}
\label{table:3}
\end{table*}
\end{landscape}

\end{document}